\documentclass[12pt]{article}

\usepackage{a4wide,amsmath,amssymb,amsthm}
\usepackage{amsfonts}
\usepackage{fix-cm}
\usepackage{mathtools}
\usepackage{extarrows}
\usepackage{enumerate}
\usepackage{nicefrac}
\usepackage{afterpage}
\usepackage[12pt]{moresize}

\usepackage[utf8]{inputenc}
\usepackage[portuguese,main=english]{babel}
\usepackage[pdftex]{graphicx}	
\usepackage[authoryear]{natbib}
\usepackage[affil-it]{authblk}

\usepackage{geometry}
\usepackage[]{booktabs}
\usepackage{multirow,multicol}
\usepackage[flushleft]{threeparttable}
\usepackage{lscape}
\usepackage{longtable}

\usepackage{complexity}
\usepackage{marginnote}
\usepackage{textcomp}

\usepackage{url}
\usepackage{color}
\usepackage[x11names]{xcolor}
\colorlet{linkequation}{blue}
\usepackage[colorlinks]{hyperref}

\newcommand*{\SavedEqref}{}
\let\SavedEqref\eqref
\renewcommand*{\eqref}[1]{
\begingroup
\hypersetup{
    linkcolor=linkequation,
    linkbordercolor=linkequation,
}
\SavedEqref{#1}%
\endgroup
}

\hypersetup{
    citecolor=SpringGreen4,
    citebordercolor=SpringGreen4,
}

\allowdisplaybreaks
\usepackage{setspace}
\theoremstyle{definition}

\theoremstyle{remark}

\newcommand{\defeq}{\stackrel{\text{def}}{=}}

\title{\Large 
Polyhedral approach to weighted connected matchings in general graphs}
\author[1]{Phillippe Samer}
\author[2]{Phablo F.S. Moura}
\affil[ ]{\small 
{\tt $\lbrace$samer@uib.no, phablo.moura@kuleuven.be$\rbrace$ }
}
\affil[1]{\small 
Universitetet i Bergen, 
Institutt for Informatikk, 
Postboks 7800, 5020, Bergen, Norway
}
\affil[2]{\small 
Research Center for Operations Research \& Statistics, KU Leuven, Belgium
}

\date{October 7, 2023}

\begin{document}

\maketitle

\begin{abstract}

A connected matching in a graph $G$ consists of a set of pairwise disjoint edges whose covered vertices induce a connected subgraph of $G$.
While finding a connected matching of maximum cardinality is a well-solved problem, it is NP-hard to determine an optimal connected matching in an edge-weighted graph, even in the planar bipartite case.
We present two mixed integer programming formulations and a sophisticated branch-and-cut scheme to find weighted connected matchings in general graphs.
The formulations explore different polyhedra associated to this problem, including strong valid inequalities both from the matching polytope and from the connected subgraph polytope.
We conjecture that one attains a tight approximation of the convex hull of connected matchings using our strongest formulation, and report encouraging computational results over DIMACS Implementation Challenge benchmark instances.
The source code of the complete implementation is also made available.

\

\begin{footnotesize}
\textbf{Acknowledgements}.
P.Samer gratefully acknowledges the work of the institute administration at UiB, according to the Working Environment Act~\textsection14-7 of the Royal Norwegian Ministry of Labour and Social Inclusion, which enabled the mobility period leading to the research results presented in this paper, as well as the support by the Research Council of Norway through the research project 249994~CLASSIS.
\end{footnotesize}
\end{abstract}

\newpage


\section{Introduction}
\label{sec:intro}

A \textsf{P}-matching in a graph $G$ consists of a matching $M$ such that the subgraph induced by vertices covered by $M$ has some property \textsf{P}, \textit{e.g.} being connected.
This paper is devoted to the problem of computing maximum weight \textit{connected matchings} in a general graph: a set of pairwise disjoint edges of maximum total weight, whose covered vertices induce a connected subgraph of $G$.

Exciting results on the computational tractability of determining connected matchings attract justified attention in the literature around this appealing generalization of classical matchings, which are already such a fundamental structure in bridging theory and sophisticated applications across domains that range from early logistics, as the postperson problem illustrates, to novel programs in kidney paired exchange \citep{Lam2020}.
A striking dichotomy here is that finding a maximum cardinality connected matching is a well-solved problem, while the edge-weighted counterpart is NP-hard even in very restricted graph classes.
As we outline below, our contributions represent a step forward in sharpening our ability to face that challenge, proposing a polyhedral combinatorics framework to actually determine maximum weight connected matchings in practice.

We remark that work on \textsf{P}-matching problems dates back at least to \cite{StockmeyerVazirani1982} on induced matchings.
Increased attention is due to thorough advances by \cite{golumbic2001uniquely} on uniquely restricted matchings, and \cite{Goddard2005disconnected} contemplating acyclic, connected, and disconnected matchings.
More recently, a number of fine-grained complexity results about the weighted connected matching (WCM) problem were presented by \cite{gg2022latin,gg2023tcs}.
They establish the NP-hardness of finding maximum weight connected matchings, for instance, even in planar graphs of maximum vertex degree 3 and edge weights in $\left\lbrace -1, 1 \right\rbrace$, and in planar bipartite graphs with edge weights in $\left\lbrace 0, 1 \right\rbrace$.

In light of that complexity barrier, our hope is to bring the machinery of polyhedral studies and mixed-integer linear programming (MILP) to bear on the investigation of WCM in general graphs.
We seek to contribute both to the theoretical study in understanding and approximating the polytope $\mathfrak{C}(G)$ of connected matchings  (\textit{i.e.} the convex hull of characteristic vectors in $\mathbb{Re}^{|E(G)|}$ of connected matchings in $G$), and to practical algorithms and their computer implementation.
We now stand on decades' worth of progress in matching theory, in combinatorial optimization problems around connectivity and network design, and in mathematical programming computation.
It is therefore expected that we are able to harness the polyhedral point of view, and evaluate to what extent it leads to the practical solution of the WCM problem.

The main idea in this paper is that there are powerful, elegant polyhedral descriptions of WCM in general graphs, in the sense that we may expect a strong foundation of polyhedral results and progressively more effective MILP solvers for this problem.
We defend the standpoint that only the combination of theoretical and applied results from communities in combinatorics and mathematical programming may truly settle (and push) the limitations around finding optimal weighted connected matchings.
From this perspective, the carefully designed formulations and the open-source software that we propose are useful ingredients towards that end.

In summary, our main contributions are the following.
\begin{enumerate}

\item Polyhedral descriptions yielding exact integer programming algorithms to find weighted connected matchings in general graphs. We present both a compact, extended formulation that can be easily fed to a black-box solver, and an exponential formulation on the space of natural variables only, using blossom inequalities from the matching polytope, and minimal separators and indegree inequalities from the connected subgraph polytope.

\item Detailed presentation of a sophisticated branch-and-cut scheme based on the exponential formulation. 
The resulting algorithm, as well as the solver based on the compact formulation, attain encouraging computational performance on four different sets of benchmark instances of connected subgraph problems from the 11th DIMACS Implementation Challenge, and settles a state-of-the-art baseline for future work.

\item Free, open-souce implementation of the complete algorithms, including a series of useful, general-purpose algorithmic components -- all of the separation procedures more prominently.
\end{enumerate}


\section{Polyhedral descriptions of weighted connected matchings}
\label{sec:formulations}

In this section we present the main idea of the paper.
We concentrate here on the polyhedra leading to MILP formulations for WCM.
Section~\ref{sec:bnc} continues with algorithmic aspects, including our particular design choices based on preliminary computational evaluations.

Our terminology and notation are standard in algorithmics and graph theory.
Note that we write $[k] \defeq \left\lbrace 1, \ldots, k \right\rbrace$, and that we denote by $2^S$ the power set of $S$, that is, the set of all subsets of $S$.

\subsection{Extended formulation}
\label{sec:formulations:extended}

We begin with a compact, extended formulation.
That is, a system of inequalities in higher dimensional space which (i) has a number of variables and constraints that is polynomial in the input size, 
and (ii) whose orthogonal projection into the original space contains all (and only those) lattice points corresponding to integer feasible solutions.
In particular, we use the well-known approach of modelling the flow of a commodity in an auxiliary network to impose the connectivity of the induced  subgraph; see \cite{MagnantiWolsey1995} for a thorough introduction.

We denote the flow network by
$\mathcal{D}=(V(G) \cup \left\lbrace s \right\rbrace, \mathcal{A})$ 
where $s$ is an artificial source vertex, and $\mathcal{A}$ contains both orientations of each original edge in $G$, besides an arc from $s$ to each other vertex. That is, 
${
\mathcal{A} \defeq \left\lbrace (u,v), (v,u): \text{ for } \left\lbrace u,v \right\rbrace \in E(G) \right\rbrace
\cup
\left\lbrace (s,u): u \in V(G) \right\rbrace
}$.
As usual, let $n \defeq |V(G)|$, $m \defeq |E(G)|$, $\delta: V(G) \rightarrow 2^E$ denote the set of edges incident to a vertex of graph $G$, and let
$\delta^+, \delta^-: V(\mathcal{D}) \rightarrow 2^\mathcal{A}$ denote the set of arcs leaving (resp. entering) a vertex of network $\mathcal{D}$.

The model we propose imposes connectivity of the solution by requiring that there be an arborescence rooted in $s$, so that there is an arc reaching a given vertex if and only if it is saturated by a matching edge.
To accomplish that, we note that each matching $M$ covers $2\cdot |M|$ vertices, and so determine that exactly $2\cdot |M|$ \textit{units of flow} leave the artificial source $s$, and that each covered vertex \textit{absorbs} a flow unit.
Specifically, a first MILP formulation of the WCM problem  is given by
\begin{equation}
\max \left\lbrace \sum_{e \in E(G)} w_e x_e: 
(\mathbf{x}, \mathbf{y}, \mathbf{f})
\in 
\mathcal{P}_{\text{ext}}(G) 
\cap
\left\lbrace 0,1 \right\rbrace^{m} \times
\left\lbrace 0,1 \right\rbrace^{2m+n} \times
\mathbb{Q}_+^{2m+n}
\right\rbrace,
\label{eq:compact:obj}
\end{equation}
where $\mathcal{P}_{\text{ext}}(G)$ is the following polyhedral region:
\vspace{-0.01cm}
\begin{alignat}{2}
\sum_{e \in \delta(u)} x_{e}  & \leq  1  \hspace{0.2cm}  && \text{ for each } u \in V(G), \label{eq:compact:cnstr1}\\
\sum_{a \in \delta^{\text{--}}(u)} y_{a}  & =  \sum_{e \in \delta(u)} x_{e}  \hspace{0.2cm}  && \text{ for each } u \in V(G), \label{eq:compact:cnstr2}\\
\sum_{u \in V(G)} y_{su} & \leq 1, \label{eq:compact:cnstr3} &&  \\
y_{uv}  & \leq  \sum_{a \in \delta^{\text{--}}(u)} y_{a}  \hspace{0.2cm}  && \text{ for each } u \in V(G) \text{ and each } uv \in \delta^+(u), \label{eq:compact:cnstr4}\\
f_{a}  & \leq  n \cdot y_a  \hspace{0.2cm}  && \text{ for each } a \in \mathcal{A}, \label{eq:compact:cnstr5}\\
\sum_{u \in V(G)} f_{su}  & =  2 \cdot \sum_{e \in E(G)} x_{e}, \label{eq:compact:cnstr6} &&  \\
\sum_{a \in \delta^{\text{--}}(u)} f_{a}  -  \sum_{a \in \delta^{\text{+}}(u)} f_{a}  & =  \sum_{a \in \delta^{\text{--}}(u)} y_{a}  \hspace{0.2cm}  && \text{ for each } u \in V(G), \label{eq:compact:cnstr7}\\
x_{e} & \geq 0 \hspace{1cm}  &&  \text{ for each } e \in E(G), \label{eq:compact:bounds1} \\
y_{a} & \geq 0 \hspace{1cm}  &&  \text{ for each } a \in \mathcal{A} , \label{eq:compact:bounds2} \\
f_{a} & \geq 0 \hspace{1cm}  &&  \text{ for each } a \in \mathcal{A} . \label{eq:compact:bounds3}
\end{alignat}

Note that variables $\mathbf{x}$ determine which edges of $G$ are in the solution matching, variables $\mathbf{y}$ determine an orientation by allowing arcs of $\mathcal{D}$ to carry flow, and variables $\mathbf{f}$ give the actual flow running on each arc.

The classical degree inequalities in~$\eqref{eq:compact:cnstr1}$ are enough to have integer points where at most one edge reaches each vertex.
Constraints~$\eqref{eq:compact:cnstr2}$ link $\mathbf{x}$ and $\mathbf{y}$ variables, by setting the number of (directed) arcs entering $u$ as the number of (undirected) edges incident to it -- either zero or one, as enforced by the previous constraints.
Constraint~$\eqref{eq:compact:cnstr3}$ establishes that the artificial flow source $s$ should be linked to at most one vertex in $G$; note that we expect exactly one arc leaving $s$ in interesting examples, but the model still allows an empty solution (\textit{e.g.} when the objective coefficients $\mathbf{w}$ are negative everywhere).
Constraints~$\eqref{eq:compact:cnstr4}$, which capture that we may only open an out arc from $u$ to $v$ if some in arc arrives at $u$, are actually implied by the other sets of inequalities but generally perceived as helping solve LP relaxation faster.

The remaining constraints concern the network flow. Inequalities~$\eqref{eq:compact:cnstr5}$ bind $\mathbf{y}$ and $\mathbf{f}$ variables: nonzero flow is only allowed in open arcs, and the maximum flow is $n = |V(G)|$.
Constraint~$\eqref{eq:compact:cnstr6}$ establishes that the flow leaving the artificial source $s$ is exactly the number of vertices saturated by the matching (\textit{i.e.} twice as many as there are edges in the matching).
Lastly, flow balance constraints~$\eqref{eq:compact:cnstr7}$ impose a single connected component in the solution: vertices in the arborescence (namely, those whose number of incoming arcs in the right-hand side is one) consume one unit of flow, while others may not interfere either consuming or creating flow.

The main advantage of having a MILP formulation with polynomial number of variables and constraints is the practicality of just feeding it to a black-box solver, automatically benefiting of increased performance due to software and hardware improvement.
On the other hand, while an extended formulation may have much smaller number of facets than its projection, decades of mathematical programming computation led to numerous examples where superior performance is attained by branch-and-cut algorithms that dynamically identify and add cutting-planes violated by relaxation solutions.
That is the path we now thread. First, designing a formulation for WCM anchored in solid knowledge of the underlying connected subgraph polytope and the classical matching polytope.
Next, in Section~\ref{sec:bnc}, filling in the details of the best-performing branch-and-cut scheme we devised and offer in our accompanying software repository.

\subsection{Formulation in the original space of variables}
\label{sec:formulations:separators}

The guiding principle in the design of our second formulation for WCM is to waive the concern about model size and build on strong valid inequalities leading to the best-performing solvers for closely related problems, defined over the larger polytopes of classical matchings and connected subgraphs.

The classical matching polytope is well-known since the very birth of the polyhedral combinatorics field -- tied to the celebrated results of \cite{edmonds1965maximum}, and better understood in light of the combinatorial proof of \cite{balinski1972establishing}.
Namely, Edmonds showed that, together with degree inequalities in $\eqref{eq:compact:cnstr1}$ above, \textit{blossom inequalities} give all the facets missing in a complete characterization of the matching polytope, and can be separated efficiently.

On the other hand, as the maximum-weight connected subgraph (MWCS) problem is \textsf{NP}-hard even in very restricted particular cases~\citep{johnson1985np}, there is no hope for an ideal formulation of the connected subgraph polytope under the assumption that $\textsf{P}\neq \textsf{NP}$ and the equivalence of separation and optimization.  
While there are many options for modelling induced connectivity, a recent performance breakthrough in exact solvers for problems like the MWCS \citep{wang2017imposing,alvarez2013maximum} and Steiner trees \citep{fischetti2017MPC} is attributed to \textit{minimal separator inequalities} (MSI) on vertex choosing variables $y \in \left\lbrace 0,1 \right\rbrace^{|V(G)|}$:
\begin{equation}
\label{eq:MSI}
y_a + y_b -\sum_{u \in S} y_u \leq 1,
\end{equation}
for each pair of non-adjacent vertices $a$ and $b$, and each $(a,b)$-separator $S \subseteq V \backslash \left\lbrace a,b \right\rbrace$, \textit{i.e.} there are no paths connecting $a$ to $b$ if we remove $S$ from $G$.
The eminently readable paper by \cite{wang2017imposing} includes a proof that $\eqref{eq:MSI}$ defines a facet of the connected subgraph polytope if and only if the separator $S$ is minimal.

It is worth remarking that the breakthrough we refer to in practical evidence (runtime of resulting MILP solvers) does not agree with the theoretical intuition given by inclusion of different polyhedral relaxations.
In particular, the recent work of \cite{rehfeldt2022optimal} proves that the LP bound from the MSI relaxation for induced connectivity is weaker than earlier alternatives based on combining vertex and edge-variables.
That is in stark contrast to the experimental results mentioned above.
In particular, the praised solver of \cite{fischetti2017MPC} won most of the categories at the 11th DIMACS Implementation Challenge \citep{dimacs11website}.
In line with our guiding principle in this section, we follow the standpoint of those authors, who conclude that a simpler model defined in the natural space of variables and a careful implementation framework enabled their superior performance.

To define a system of inequalities combining the separation of MSI $\eqref{eq:MSI}$ and only using natural design variables $\mathbf{x} \in \left\lbrace 0,1 \right\rbrace^{|E(G)|}$ in the original space of polytope $\mathfrak{C}(G)$,
we use the fact that vertex $u$ belongs to the subgraph induced by matching $M$ if and only if there is exactly one edge in $M$ incident to $u$.
Hence, projecting the MSI onto the space of $\mathbf{x}$ variables using
$y_u \mapsto \sum_{e \in \delta(u)} x_e$, 
we derive the first IP formulation to find maximum weight connected matchings using MSI,
\begin{equation}
\max \left\lbrace \sum_{e \in E(G)} w_e x_e: 
\mathbf{x}
\in 
\mathcal{P}_{\text{sep}}(G) 
\cap
\left\lbrace 0,1 \right\rbrace^{m}
\right\rbrace,
\label{eq:msi_only:obj}
\end{equation}
with $\mathcal{P}_{\text{sep}}(G)$ defined by degree inequalities $\eqref{eq:compact:cnstr1}$, non-negativity bounds $\eqref{eq:compact:bounds1}$,  and the projection of separator inequalities $\eqref{eq:MSI}$:
\vspace{-0.01cm}
\begin{small}
\begin{alignat}{2}
\sum_{e \in \delta(a)} x_e + \sum_{e \in \delta(b)} x_e -\sum_{u \in S} \sum_{e \in \delta(u)} x_e & \leq 1  \hspace{0.5cm} && \text{ for each } a,b \in V(G), \left\lbrace a,b\right\rbrace \not\in E(G), \text{ and each } S \in \mathcal{C}(a,b), \nonumber
\end{alignat}
\end{small}
where $\mathcal{C}(a,b)\subset 2^{V(G)}$ denotes the set of all minimal $(a,b)$-separators in $G$.

We reinforce this formulation with two exponential families of valid inequalities.
The first is an additional class of facets of the connected subgraph polytope studied by \cite{wang2017imposing}.
The other was already mentioned earlier: blossom inequalities, which define all remaining facets of the classical matching polytope.

\begin{description}
\item[Indegree inequalities] 
A vector $d \in \mathbb{Z}_+^n$ is called an \textit{indegree} vector of graph $G$ if there exists an orientation of its edges such that the indegree of each vertex $u$ is $d_u$.
Introduced decades earlier in the context of greedoid optimization and only in the particular case where $G$ is a tree \citep[Chapter XI, Theorem 3.6]{korte1991greedoids}, it was later shown by \cite{wang2017imposing} that, for each indegree vector $d$ of an \textit{arbitrary} graph $G$, inequality
\[
\sum_{u \in V(G)} (1-d_u)\cdot y_u \leq 1
\]
is valid for the connected subgraph polytope of $G$.

Interestingly, the indegree inequalities provide a minimal, complete characterization of that polytope when $G$ is a tree: each indegree inequality defines a facet, and none is missing.
More importantly in the context of our problem, \cite{wang2017imposing} prove that the indegree inequalities can still define facets in the general case, and may be separated in time proportional to $O(n+m)$.
Again, we shall use the projection of those inequalities in the original space of the connected matching polytope $\mathfrak{C}(G)$ by the linear map $y_u \mapsto \sum_{e \in \delta(u)} x_e$.

\item[Blossom inequalities] 
Finally, to ensure that our formulation is within the tightest possible  description of the (classical) matching relaxation, one would naturally expect the inclusion of blossom inequalities.
Namely, for each \textit{handle} $H \subset V(G)$ of odd cardinality, the inequality
\[
\sum_{e \in E(G[H])} x_e \leq \dfrac{|H|-1}{2}
\]
is valid for the matching polytope of $G$.
Besides being sufficient to determine the convex hull of incidence vectors of matchings in an arbitrary graph,
each blossom inequality is also necessary when $G$ is a complete graph, for example.
They are also an important ingredient in state-of-the-art solvers for the travelling salesperson problem \citep[Section 7.4]{CCZ2014book}. 
\end{description}

Putting the inequalities together, the complete formulation on which we base our enhanced branch-and-cut algorithm to find weighted connected matchings is
\begin{equation}
\max \left\lbrace \sum_{e \in E(G)} w_e x_e: 
\mathbf{x}
\in 
\mathcal{P}_{\text{full}}(G) 
\cap
\left\lbrace 0,1 \right\rbrace^{m}
\right\rbrace,
\label{eq:full:obj}
\end{equation}
where $\mathcal{P}_{\text{full}}(G)$ is the following polyhedral region:
\vspace{-0.01cm}
\begin{footnotesize}
\begin{alignat}{2}
\sum_{e \in \delta(u)} x_{e}  & \leq  1  && \text{ for each } u \in V(G), \label{eq:full:degree}\\
\sum_{e \in \delta(a)} x_e + \sum_{e \in \delta(b)} x_e -\sum_{u \in S} \sum_{e \in \delta(u)} x_e & \leq 1  && \text{ for each non-adjacent } a,b \in V(G), \text{ and each } S \in \mathcal{C}(a,b),\label{eq:full:msi} \\
\sum_{u \in V(G)} (1-d_u) \sum_{e \in \delta(u)} x_e  & \leq 1  && \text{ for each indegree vector } d \text{ of } G,\label{eq:full:indegree}\\
\sum_{e \in E(G[H])} x_e &
\leq \dfrac{|H|-1}{2} \hspace{0.2cm} && \text{ for each } H \subset V(G) \text{ with } |H| \text{ odd}, \label{eq:full:blossom}\\
x_{e} & \geq 0  &&  \text{ for each } e \in E(G). \label{eq:full:bounds}
\end{alignat}
\end{footnotesize}


\section{Branch-and-cut scheme for the exponential formulation}
\label{sec:bnc}

The enhanced WCM formulation $\eqref{eq:full:obj}$ is only useful in practice if an implementation of efficient separation procedures for the (exponentially-many) inequalities in $\eqref{eq:full:msi}$, $\eqref{eq:full:indegree}$ and $\eqref{eq:full:blossom}$ defining $\mathcal{P}_{\text{full}}(G)$ is available.
This section completes our main contribution, filling in the algorithmic details and presenting our free, open-source software package with the resulting solver for WCM in general graphs.

We designed our \textsf{C++} code with attention to time and space efficiency, and
fairly tested it for correctness along months of development.
It is available in the \textsf{wcm-branch-and-cut} repository on GitHub (\url{https://github.com/phillippesamer/wcm-branch-and-cut}), thus welcoming collaboration towards extensions and facilitating the direct comparison with eventual algorithms designed for the WCM problem in the future.
Moreover, the code can be forked and turned into useful algorithmic components to different problems and applications.

\subsection{Separation procedures}
\label{sec:bnc:separation}

Efficient separation algorithms are at the core of a successful branch-and-cut scheme, as they are executed a number of times in each node of the enumeration tree partitioning the search space.
Since the classes of inequalities $\eqref{eq:full:msi}$, $\eqref{eq:full:indegree}$ and $\eqref{eq:full:blossom}$ grow exponentially with the size of the input graph, it is not practical to add them in an explicit model \textit{a priori} for reasonably sized problems.
Except for the last method presented below, the following are \textit{exact} separation procedures: oracles that, given a relaxation solution $\mathbf{x}^*$, either identify a specific inequality valid for $\mathcal{P}_{\text{full}}(G)$ that is violated at $\mathbf{x}^*$ (which can then be added to the formulation and \textit{cut off} $\mathbf{x}^*$ to continue the search), or certify implicitly that no such inequality exists.
In contrast, when it comes to blossom inequalities $\eqref{eq:full:blossom}$, we use both an exact and a \textit{heuristic} separation procedure, \textit{i.e.} a faster method to search for a blossom cut, that may fail even when one exists.

\subsubsection*{MSI cuts}

We followed the description of \cite[Section 2.1]{fischetti2017MPC} for two exact separation algorithms for MSI in the award-winning solver mentioned in Section~\ref{sec:formulations:separators}.

The first algorithm is based on the computation of maximum flows in a support digraph $H$, whose arc capacities are defined according to the current relaxation solution.
For each pair of non-adjacent vertices $u,v \in V(G)$ such that $x_u^* + x_v^* > 1$ (which is necessary for an MSI to be violated), we calculate a maximum ($u,v$)-flow $\overline{f}$ in $H$.
If $\overline{f} < x_u^* + x_v^* -1$, we may determine a violated separator inequality from the corresponding minimum cut.
Two implementation tweaks are worth mentioning.
\begin{enumerate}[(A)]
\item As first observed by \cite[Section 3.1]{phablo2021ejor}, we often have a large number of variables in the relaxation solution $\mathbf{x}^*$ at either $0$ or $1$, and none of these appear in a $(u,v)$-separator that gives a violated inequality. 
We may therefore contract the corresponding arcs and vertices in $H$ to run the separation algorithm in considerably smaller support digraphs.
Implementing such contractions requires special care to keep track of the original variables that make up the violated inequality we eventually find.

\item When we identify a minimum cut $C$ yielding a violated $(u,v)$-separator inequality, it is in general not a \textit{minimal} separator.
As mentioned in Section~\ref{sec:formulations:separators}, we thus have the opportunity to \textit{lift} the left-hand  side towards a minimal separator $S \subset C$ to derive a non-dominated inequality.
This can be achieved with a simple graph traversal procedure, as formalized by \cite[Section 2.1]{fischetti2017MPC}.
While they refer to a breadth-first search (BFS) and use an edge-deletion operation, we observe faster runtimes combining (i) a depth-first search that avoids an explicit BFS queue, and (ii) a boolean mask of \textit{active}/\textit{inactive} edges passed as a reference parameter instead of modifying the graph.
\end{enumerate}
The worst-case time complexity of the whole procedure is in $O(n^2 \cdot g(\tilde n,\tilde m))$, where $n \defeq |V(G)|$, $m \defeq |E(G)|$, and $g(\tilde n,\tilde m)$ denotes the complexity of a single maximum-flow computation in a digraph with $\tilde n$ vertices and $\tilde m$ arcs.
We use the highly tuned implementation of the preflow-push algorithm of \cite{GoldbergTarjan1988} in COIN-OR LEMON 1.3.1 -- the Library for Efficient Modeling and Optimization in Networks~\citep{lemon2011}, as the responsible team reports that best computational runtimes are attained with that algorithm.
Its time complexity $g(\tilde n,\tilde m)$ is in $O({\tilde n}^2 \sqrt{\tilde m})$.
Since the digraph $H$ above is such that $\tilde n \leq 2n$ and $\tilde m \leq 2m+n$ before the contractions due to integer-valued variables, the runtime of our separation is within $O(n^4 \sqrt{n+m})$.
Contrary to what one could expect from such a high worst-case time complexity, we had very encouraging numerical results in practice, which we attribute to the digraph contractions and the particular branch-and-cut scheme that we outline in Section~\ref{sec:bnc:general} below.

An alternative, more efficient separation procedure is readily available in the particular case where the relaxation solution $\mathbf{x}^*$ is actually integer-valued.
We may resort to a simple depth-first search (DFS) to check connectivity of the induced solution.
In a disconnected solution, inspecting the neighbourhood $C$ of any connected component with a vertex $u$ such that $x_u^* = 1$ gives a violating $(u,v)$-separator, for some $v$ in a different component with $x_v^* = 1$.
It is important to stress that implementation tweak (B) above applies here as well, and that we still need to derive a minimal separator $S \subset C$ to add a stronger inequality.
The time complexity of the separation in this particular case is dominated by the DFS step, and is thus in $O(n + m)$.

It is fair to remark that MSI first appeared in two, earlier papers in applied operations research.
\cite{fugenschuh2008} introduced this class of inequalities and their separation algorithm in an intricate, non-linear programming problem arising in the sheet metal industry.
Their work was picked up by \cite{carvajal2013}, who extended on the role of MSI when imposing connectivity in a forestry planning problem.
The MSI were also introduced independently in the polyhedral studies of the convex recoloring problem \citep{campelo2013CR}.
We also praise, again, the thorough, instructive chapter of \cite{alvarez2013maximum} on the maximum weight connected subgraph problem more generally.

\subsubsection*{Indegree cuts}

The separation of indegree inequalities is remarkably simple.
We implemented the procedure exactly as presented by \cite{wang2017imposing}.
The main point is to consider an orientation maximizing the left-hand side over all indegree vectors, namely: taking $\overrightarrow{uv}$ for $\left\lbrace u,v \right\rbrace \in E(G)$ if and only if $x_u^* \geq x_v^*$.
The worst-case time complexity of the algorithm is in $O(n+m)$.

\subsubsection*{Blossom cuts}

We consider two separation algorithms for blossom inequalities. See Section~\ref{sec:bnc:general} below for the detailed scheme in which they are used.

The first method is the exact separation procedure of \cite{PadbergRao1982}. 
We strictly followed its presentation \textit{cf.} \cite{letchford2008blossom}, who introduced the state-of-the-art algorithm for the more general version of $b$-matchings with edge capacities.
The separation works on a support graph $G^\prime$ with $n+1$ vertices and $m+n$ edges, whose capacities are determined by the current relaxation solution $\mathbf{x}^*$.
It boils down to determining a \textit{cut tree} $\mathcal{T}(G^\prime)$: an elegant data structure introduced by \cite{GomoryHu1961} to encode minimum cuts between all pairs of vertices in the graph, at the expense of computing only $|V(G^\prime)|-1$ maximum flow computations.

We use COIN-OR LEMON 1.3.1 \citep{lemon2011} here as well to build the cut tree $\mathcal{T}(G^\prime)$.
Then, it suffices to verify the blossom inequality $\sum_{e \in E(G[H])} x_e^* \leq \nicefrac{(|H|-1)}{2}$ for at most one handle $H \subset V(G)$ per edge of the tree.
Constructing the data structure dominates the worst-case time complexity of the complete algorithm, and is within $O(n^4)$ \citep{letchford2008blossom}.
We remark that an implementation here may construct the support graph $G^\prime$ only once, and just update edge capacities according to  the current relaxation values.

Our second method is inspired by the work of \cite[Section 4.1.2]{bicalho2016bcp}, who observed comparable results using the exact method above and a separation heuristic for blossom inequalities in a row-and-column generation algorithm for a different network design problem.
We devised a simpler algorithm to try to find blossom cuts in linear time as follows.
Let $H$ denote the support graph induced only from fractional variable in $\mathbf{x}^*$, and let $H_i$ denote the connected components of $H$.
For each $H_i$ of odd cardinality, we simply inspect the corresponding blossom inequality for violation.
The complexity of this separation heuristic is dominated by the step finding connected components in $H$, which is in $O(n+m)$ in the worst case.

\subsection{Further algorithmic aspects}
\label{sec:bnc:general}

We are now ready to depict our complete branch-and-cut scheme.
We use the overall framework in the Gurobi 10.0.2 solver \citep{gurobi}, with callbacks to implement separation procedures.
It is important to distinguish between the specific callback from a new MIP incumbent, where only \textit{lazy constraints} are added (in our case, MSI tailored for integer-valued points), and the standard callback from the search at a given MIP node, where we add \textit{user cuts} (all of MSI, indegree, and blossom inequalities).

In the beginning, only degree inequalities $\eqref{eq:full:degree}$ are included \textit{a priori} in the model.
Instead of relying in the solver standard behavior concerning how long to explore the root node relaxation before branching, we designed a \textbf{strengthened root node LP relaxation}.
Here, we dedicate up to 300 seconds or 10\% of the specified time limit (whichever is shorter) prior to the main call to the solver, and alternate the solution of an LP relaxation and cut generation observing that:
\begin{enumerate}[1.]
\item All MSI violated in the current relaxation solution $\mathbf{x}^*$ are added;
\item The exact separation of blossom inequalities is attempted only if $\mathbf{x}^*$ is fractional, no MSI cut was found and the separation heuristic failed;
\item No indegree cuts are added unless all other separation algorithms failed, to prevent the inclusion of an excessively large number of constraints.
\end{enumerate}

The reinforced model resulting from this initialization consistently showed the best computational performance across a range of configurations in our preliminary computational evaluation.
In particular, we explain in Section~\ref{sec:xp:results} that a number of instances could be solved without resorting to branching -- which was not the case before we devised this strengthened root node relaxation.

The general algorithm continues with the branch-and-cut enumeration tree, partitioning the search space by fixing a single binary variable, $x_u = 0$ or $x_u = 1$, in each branch.
In each remaining node below the root relaxation, more attention is paid to limit the complexity of cut generation, by observing the following:
\begin{enumerate}[1.]
\item The MSI separation algorithm concludes as soon as any inequality violated at the current relaxation is found, iterating over the initial source vertex considered for maximum flow computations;
\item The exact separation of blossom inequalities is turned off, and only the heuristic is run;
\item The separation of indegree inequalities is turned off.
\end{enumerate}


\section{Experimental evaluation}
\label{sec:xp}

We conclude our work with the first experimental evaluation of an algorithm for the weighted connected matching (WCM) problem.
The main goal here is to indicate that our polyhedral descriptions and the resulting algorithms constitute a practical approach to find optimal connected matchings in non-trivial inputs.

\subsection{Benchmark design}
\label{sec:xp:design}

We hope to set a judicious baseline towards progress in the computation of WCM. 
Namely, one that is anchored in the \textit{reproducibility} of materials and methods, and that reports experimental evidence from respectable, interesting testbeds.

\subsubsection*{Interlude} 
At an early stage of our experiments, we considered using binomial (Erd\H{o}s-R\'{e}nyi) graphs $\mathcal{G}_{n,p}$ as reported by \cite{wang2017imposing} when studying the performance of minimal separator and indegree inequalities for the maximum weight connected subgraph (MWCS) problem.
To our surprise, the random graphs in this model resulted in quite simple WCM instances. 
Both the compact extended formulation and the exponential one in the original space give so tight bounds that problems in the order of $10^5$ vertices were solved in negligible time on a desktop computer -- whether on sparse or dense $\mathcal{G}_{n,p}$~graphs ($p \in [0.01, 0.6]$), whether with Gaussian or uniform random weights.
The script to produce such instances using the robust \textsf{NetworkX} Python package is still available in our GitHub repository.
Nevertheless, we do not go further in evaluating those examples in our research.
Instead, we choose to proceed by borrowing credibility from a certified source of benchmark instances of MWCS and similar problems.

\subsubsection*{Benchmark instances}
Our computational evaluation is carried over benchmark instances from the 11th DIMACS/ICERM Implementation Challenge. 
The competition covered several variants around the Steiner tree problem, and we chose to use all three sets of instances of the MWCS problem, and the one set available for the Generalized Maximum-Weight Connected Subgraph (GMWCS) problem.
Specifically, there are
39 instances in set \textsf{MWCS-GAM}, 
72 in \textsf{MWCS-JMPALMK}, 
8  in \textsf{MWCS-ACTMOD}, and 
63 instances in set \textsf{GMWCS-GAM}.
Table~\ref{tab:instances} in the supplementary material (Appendix~\ref{sec:appendix}) contains the full names, sizes, and a numerical ID for ease of reference of the 182 instances.
The smallest instances have less than a thousand vertices and edges; the largest ones exceed 5.000 vertices and 90.000 edges.
See \cite{dimacs11website} for more information.

Since MWCS instances contain only vertex weights, we determined $w(e) \defeq w(u) + w(v)$ for each $e = \left\lbrace u,v \right\rbrace \in E(G)$.
For GMWCS problems we used the edge weights included in the input instance, and ignored vertex weights.
We note that 10 out of the 63 GMWCS intances have only negative weights, and so the resulting WCM problem has null optimum.
We decided to keep those instances in the benchmark for the sake of completeness.
%
%

\subsubsection*{Platform and settings}
We tested the implementation in a desktop machine with an Intel Core i5-8400 processor, with 6~CPU cores at 2.80GHz, and 16GB of RAM, runnning GNU/Linux kernel 6.2.0-33 under the U\-bun\-tu~22.04.3 LTS distribution.
All the code is compiled with g++ 11.4.0.

As mentioned in Section~\ref{sec:bnc:general}, we use the \cite{gurobi} MILP solver.
We set a time limit of 3600 seconds in all executions, while noting that the solver process may exceed that by a negligible amount.
All solver parameters are used in their default values, except for   setting an extra effort on MIP heuristics when using the exponential formulation with our branch-and-cut scheme ({\footnotesize \texttt{MIPFocus = 1} and \texttt{GRB\_DoubleParam\_Heuristics = 0.2}}).
In our implementation, we require a violation by at least $10^{-5}$ to add a cutting plane in all separation procedures.

\subsection{Discussion}
\label{sec:xp:results}

Table~\ref{tab:xp} in the supplementary material in Appendix~\ref{sec:appendix} contains the detailed results of the solver using the extended formulation $(\ref{eq:compact:obj})$,  \textit{i.e.} optimizing over polyhedron $\mathcal{P}_{\text{ext}}$, and our enhanced branch-and-cut scheme with formulation $(\ref{eq:full:obj})$, which is based on the exponential polyhedral description in $\mathcal{P}_{\text{full}}$.
We include information both on the LP relaxation and the full integer program on the table, and discuss our findings next.

\paragraph{Overall performance.}
A first note is about the actual practicality of the implementations.
We were satisfied that the enhanced branch-and-cut scheme over $\mathcal{P}_{\text{full}}$ concludes with an optimality certificate for 168 out of 182 instances.
The corresponding number for the compact formulation is 151 cases.
Taking into account that we use at most one hour of processing by a regular desktop computer, we consider this a rather encouraging conclusion.
Practitioners and applications with a connected matching subproblem should therefore be able to derive improved runtimes by taking advantage of more powerful computing platforms.

\paragraph{Empirical approximation of the ideal formulation.}
Concerning how tightly we approximate the convex hull $\mathfrak{C}(G)$ with our formulations, it is remarkable that the LP relaxation bound of $\mathcal{P}_{\text{full}}$ matches the optimum in 98 out of 168 instances for which the optimum is known, and the enhanced branch-and-cut algorithm is able to prove optimality in the root node relaxation in 111 problems.
More generally, the optimum is within $5\%$ of LP bound in 145 of those 168 instances.
We observe that the LP relaxation bound of $\mathcal{P}_{\text{full}}$ is stronger or equal to that of $\mathcal{P}_{\text{ext}}$ in all instances. 
It is $23.9\%$ tighter on average, and up to $84\%$ tighter (recall that we impose a time limit for the LP relaxation, so it could be even stronger).

\paragraph{Comparing bounds between formulations.}
As expected, the dual bounds attained with the enhanced branch-and-cut algorithm over $\mathcal{P}_{\text{full}}$ are consistently stronger than that of the compact extended formulation, and there is only a single case where the latter is stronger (namely, the instance of ID 16).
Concerning primal bounds, we were surprised positively with 8 examples where the compact extended formulation does find an integer feasible solution better than the exponential one. 
We refer to the results on instances with identifiers $3, 16, 21, 35, 36, 42, 44, 45$.

\paragraph{Superiority of exponential formulation in harder instances.}
Finally, seeking a classification of the instances with respect to computational hardness, we find that 123 out of 182 instances could be solved to optimality by both formulations within 5 minutes.
In the remaining 59 instances, we find good clues of the superiority of $\mathcal{P}_{\text{full}}$.
\begin{enumerate}[i.]
\item In 30 of those 59 cases, the exponential formulation does finish within 5 minutes.
\item The LP relaxation bound of $\mathcal{P}_{\text{full}}$ is up to $65.0\%$ stronger than that of $\mathcal{P}_{\text{ext}}$ in this subset; it is $34.7\%$ stronger on average.
\item In 11 instances, the exponential formulation solves the problem at the root relaxation node, whereas the compact one struggles to finish: not even proving optimality before 3600 seconds in 3 cases; taking 1510 or 2388 seconds in 2 cases; and 6 other cases taking longer than 5 minutes.

\item While there are 14 instances where the solver with the exponential formulation could not prove optimality (\textit{i.e.} exceeds the one hour time limit with an open gap), there are 31 such cases for the compact formulation.
\end{enumerate}


\section{Final remarks}
\label{sec:conclusion}

The standing argument behind our work is that polyhedra and MILP should lead to an interesting, useful perspective to study WCM both in theory and in practice. 
This complements other methodologies that are currently available to find weighted connected matchings, which assume restricted input graph classes.
Moreover, we hope that our approach also determines a solid baseline of comparison for further research crafting WCM algorithms.

Besides their appealing polyhedral structures and intrinsic connections with established problems in combinatorics and optimization, both formulations considered here had encouraging computational performance, and could be considered for eventual applications of the WCM problem as well.
On the one hand, having good results from the compact extended formulation is an achievement in its own right (we remark that it was able to find better primal feasible solutions in a few examples), as performance improvements in the underlying MILP solver usually leads to better runtimes in such ``simpler'' models automatically.

Still, all the work in designing an enhanced formulation did pay off, and we are proud to contribute yet another success story where the theoretical insight gathered from careful polyhedral relaxations translates to strides in practical computing experience.
Using the exponential description and the resulting branch-and-cut scheme, we provide optimality certificates for 168 out of 182 instances.
Most noticeable, that formulation solves 111 out of 182 instances in the root relaxation, without resorting to branching.

We believe that further research should consider only the subset of harder instances discussed here, and investigate features that characterize the most challenging ones before proposing new benchmark sets.
It should also be possible to strengthen the compact extended polyhedron as well, so that more instances could be solved to proven optimality within limited runtimes.

Finally, our software repository includes not only the implementation of all the methods presented in this paper, but also a simple tool using \textsf{polymake} \citep{polymake2000,polymake2017} to assist one in inspecting the connected matching polytope and finding new classes of strong valid inequalities.
We had some progress in this direction \citep{samer2023}, and trust that many fruitful results could be derived by further research translating the polyhedral insight to improved WCM algorithms.



\begin{thebibliography}{31}
\providecommand{\natexlab}[1]{#1}
\providecommand{\url}[1]{\texttt{#1}}
\expandafter\ifx\csname urlstyle\endcsname\relax
  \providecommand{\doi}[1]{doi: #1}\else
  \providecommand{\doi}{doi: \begingroup \urlstyle{rm}\Url}\fi

\bibitem[{\'A}lvarez-Miranda et~al.(2013){\'A}lvarez-Miranda, Ljubi{\'{c}}, and
  Mutzel]{alvarez2013maximum}
E.~{\'A}lvarez-Miranda, I.~Ljubi{\'{c}}, and P.~Mutzel.
\newblock The maximum weight connected subgraph problem.
\newblock In M.~J{\"u}nger and G.~Reinelt, editors, \emph{Facets of
  Combinatorial Optimization: Festschrift for Martin Gr{\"o}tschel}, pages
  245--270. Springer Berlin Heidelberg, 2013.
\newblock URL \url{https://doi.org/10.1007/978-3-642-38189-8_11}.

\bibitem[Assarf et~al.(2017)Assarf, Gawrilow, Herr, Joswig, Lorenz, Paffenholz,
  and Rehn]{polymake2017}
B.~Assarf, E.~Gawrilow, K.~Herr, M.~Joswig, B.~Lorenz, A.~Paffenholz, and
  T.~Rehn.
\newblock Computing convex hulls and counting integer points with polymake.
\newblock \emph{Mathematical Programming Computation}, 9:\penalty0 1--38, 2017.
\newblock URL \url{https://doi.org/10.1007/s12532-016-0104-z}.

\bibitem[Balinski(1972)]{balinski1972establishing}
M.~L. Balinski.
\newblock Establishing the matching polytope.
\newblock \emph{Journal of Combinatorial Theory, Series B}, 13\penalty0
  (1):\penalty0 1--13, 1972.
\newblock URL \url{https://doi.org/10.1016/0095-8956(72)90002-0}.

\bibitem[Bicalho et~al.(2016)Bicalho, Da~Cunha, and Lucena]{bicalho2016bcp}
L.~H. Bicalho, A.~S. Da~Cunha, and A.~Lucena.
\newblock Branch-and-cut-and-price algorithms for the degree constrained
  minimum spanning tree problem.
\newblock \emph{Computational Optimization and Applications}, 63:\penalty0
  755--792, 2016.
\newblock URL \url{https://doi.org/10.1007/s10589-015-9788-7}.

\bibitem[Campêlo et~al.(2013)Campêlo, Lima, Moura, and
  Wakabayashi]{campelo2013CR}
M.~Campêlo, K.~R. Lima, P.~F. Moura, and Y.~Wakabayashi.
\newblock Polyhedral studies on the convex recoloring problem.
\newblock \emph{Electronic Notes in Discrete Mathematics}, 44:\penalty0
  233--238, 2013.
\newblock ISSN 1571-0653.
\newblock URL \url{https://doi.org/10.1016/j.endm.2013.10.036}.

\bibitem[Carvajal et~al.(2013)Carvajal, Constantino, Goycoolea, Vielma, and
  Weintraub]{carvajal2013}
R.~Carvajal, M.~Constantino, M.~Goycoolea, J.~P. Vielma, and A.~Weintraub.
\newblock Imposing connectivity constraints in forest planning models.
\newblock \emph{Operations Research}, 61\penalty0 (4):\penalty0 824--836, 2013.
\newblock URL \url{https://doi.org/10.1287/opre.2013.1183}.

\bibitem[Conforti et~al.(2014)Conforti, Cornu\'{e}jols, and
  Zambelli]{CCZ2014book}
M.~Conforti, G.~Cornu\'{e}jols, and G.~Zambelli.
\newblock \emph{Integer Programming}.
\newblock Springer Cham, 2014.
\newblock URL \url{https://doi.org/10.1007/978-3-319-11008-0}.

\bibitem[Dezs{\H{o}} et~al.(2011)Dezs{\H{o}}, J\"{u}ttner, and
  Kov\'{a}cs]{lemon2011}
B.~Dezs{\H{o}}, A.~J\"{u}ttner, and P.~Kov\'{a}cs.
\newblock {LEMON -- an Open Source C++ Graph Template Library}.
\newblock \emph{Electronic Notes in Theoretical Computer Science}, 264\penalty0
  (5):\penalty0 23 -- 45, 2011.
\newblock URL \url{https://doi.org/10.1016/j.entcs.2011.06.003}.

\bibitem[DIMACS'11()]{dimacs11website}
DIMACS'11.
\newblock {The Eleventh DIMACS Implementation Challenge in Collaboration with
  ICERM: Steiner Tree Problems}, 2014.
\newblock URL \url{https://dimacs11.zib.de/downloads.html}.
\newblock {Organized by David S. Johnson, Thorsten Koch, Renato F. Werneck, and
  Martin Zachariasen.}

\bibitem[Edmonds(1965)]{edmonds1965maximum}
J.~Edmonds.
\newblock Maximum matching and a polyhedron with 0, 1-vertices.
\newblock \emph{Journal of research of the National Bureau of Standards B},
  69\penalty0 (125-130):\penalty0 55--56, 1965.
\newblock URL \url{https://doi.org/10.6028/jres.069B.013}.

\bibitem[Fischetti et~al.(2017)Fischetti, Leitner, Ljubi{\'c}, Luipersbeck,
  Monaci, Resch, Salvagnin, and Sinnl]{fischetti2017MPC}
M.~Fischetti, M.~Leitner, I.~Ljubi{\'c}, M.~Luipersbeck, M.~Monaci, M.~Resch,
  D.~Salvagnin, and M.~Sinnl.
\newblock Thinning out steiner trees: a node-based model for uniform edge
  costs.
\newblock \emph{Mathematical Programming Computation}, 9\penalty0 (2):\penalty0
  203--229, 2017.
\newblock URL \url{https://doi.org/10.1007/s12532-016-0111-0}.

\bibitem[F{\"u}genschuh and F{\"u}genschuh(2008)]{fugenschuh2008}
A.~F{\"u}genschuh and M.~F{\"u}genschuh.
\newblock Integer linear programming models for topology optimization in sheet
  metal design.
\newblock \emph{Mathematical Methods of Operations Research}, 68:\penalty0
  313--331, 2008.
\newblock URL \url{https://doi.org/10.1007/s00186-008-0223-z}.

\bibitem[Gawrilow and Joswig(2000)]{polymake2000}
E.~Gawrilow and M.~Joswig.
\newblock Polymake: a framework for analyzing convex polytopes.
\newblock In G.~Kalai and G.~M. Ziegler, editors, \emph{{Polytopes --
  combinatorics and computation}}, pages 43--73. Springer, 2000.
\newblock URL \url{https://doi.org/10.1007/978-3-0348-8438-9_2}.

\bibitem[Goddard et~al.(2005)Goddard, Hedetniemi, Hedetniemi, and
  Laskar]{Goddard2005disconnected}
W.~Goddard, S.~M. Hedetniemi, S.~T. Hedetniemi, and R.~Laskar.
\newblock Generalized subgraph-restricted matchings in graphs.
\newblock \emph{Discrete Mathematics}, 293\penalty0 (1):\penalty0 129--138,
  2005.
\newblock ISSN 0012-365X.
\newblock URL \url{https://doi.org/10.1016/j.disc.2004.08.027}.
\newblock 19th British Combinatorial Conference.

\bibitem[Goldberg and Tarjan(1988)]{GoldbergTarjan1988}
A.~V. Goldberg and R.~E. Tarjan.
\newblock A new approach to the maximum-flow problem.
\newblock \emph{Journal of the ACM (JACM)}, 35\penalty0 (4):\penalty0 921--940,
  1988.
\newblock URL \url{https://doi.org/10.1145/48014.61051}.

\bibitem[Golumbic et~al.(2001)Golumbic, Hirst, and
  Lewenstein]{golumbic2001uniquely}
M.~C. Golumbic, T.~Hirst, and M.~Lewenstein.
\newblock Uniquely restricted matchings.
\newblock \emph{Algorithmica}, 31:\penalty0 139--154, 2001.
\newblock URL \url{https://doi.org/10.1007/s00453-001-0004-z}.

\bibitem[Gomes et~al.(2022)Gomes, Masquio, Pinto, Santos, and
  Szwarcfiter]{gg2022latin}
G.~C.~M. Gomes, B.~P. Masquio, P.~E.~D. Pinto, V.~F.~d. Santos, and J.~L.
  Szwarcfiter.
\newblock Weighted connected matchings.
\newblock In A.~Casta{\~{n}}eda and F.~Rodr{\'i}guez-Henr{\'i}quez, editors,
  \emph{LATIN 2022: Theoretical Informatics}, pages 54--70, Cham, 2022.
  Springer International Publishing.
\newblock ISBN 978-3-031-20624-5.
\newblock URL \url{https://doi.org/10.1007/978-3-031-20624-5_4}.

\bibitem[Gomes et~al.(2023)Gomes, Masquio, Pinto, {dos Santos}, and
  Szwarcfiter]{gg2023tcs}
G.~C. Gomes, B.~P. Masquio, P.~E. Pinto, V.~F. {dos Santos}, and J.~L.
  Szwarcfiter.
\newblock Disconnected matchings.
\newblock \emph{Theoretical Computer Science}, 956:\penalty0 113821, 2023.
\newblock ISSN 0304-3975.
\newblock URL \url{https://doi.org/10.1016/j.tcs.2023.113821}.

\bibitem[Gomory and Hu(1961)]{GomoryHu1961}
R.~E. Gomory and T.~C. Hu.
\newblock Multi-terminal network flows.
\newblock \emph{Journal of the Society for Industrial and Applied Mathematics},
  9\penalty0 (4):\penalty0 551--570, 1961.
\newblock URL \url{https://doi.org/10.1137/0109047}.

\bibitem[{Gurobi Optimization, LLC}(2023)]{gurobi}
{Gurobi Optimization, LLC}.
\newblock {Gurobi Optimizer Reference Manual}, 2023.
\newblock URL \url{https://www.gurobi.com}.

\bibitem[Johnson(1985)]{johnson1985np}
D.~S. Johnson.
\newblock {The NP-completeness column: an ongoing guide}.
\newblock \emph{Journal of algorithms}, 6\penalty0 (3):\penalty0 434--451,
  1985.
\newblock URL \url{https://doi.org/10.1016/0196-6774(85)90012-4}.

\bibitem[Korte et~al.(1991)Korte, Lov{\'a}sz, and Schrader]{korte1991greedoids}
B.~Korte, L.~Lov{\'a}sz, and R.~Schrader.
\newblock \emph{Greedoids}, volume~4.
\newblock Springer-Verlag Berlin Heidelberg, 1991.
\newblock URL \url{https://doi.org/10.1007/978-3-642-58191-5}.

\bibitem[Lam and Mak-Hau(2020)]{Lam2020}
E.~Lam and V.~Mak-Hau.
\newblock Branch-and-cut-and-price for the cardinality-constrained multi-cycle
  problem in kidney exchange.
\newblock \emph{Computers \& Operations Research}, 115:\penalty0 104852, 2020.
\newblock URL \url{https://doi.org/10.1016/j.cor.2019.104852}.

\bibitem[Letchford et~al.(2008)Letchford, Reinelt, and
  Theis]{letchford2008blossom}
A.~N. Letchford, G.~Reinelt, and D.~O. Theis.
\newblock Odd minimum cut sets and b-matchings revisited.
\newblock \emph{{SIAM Journal on Discrete Mathematics}}, 22\penalty0
  (4):\penalty0 1480--1487, 2008.
\newblock URL \url{https://doi.org/10.1137/060664793}.

\bibitem[Magnanti and Wolsey(1995)]{MagnantiWolsey1995}
T.~L. Magnanti and L.~A. Wolsey.
\newblock Optimal trees.
\newblock \emph{{Handbooks in operations research and management science}},
  7:\penalty0 503--615, 1995.
\newblock URL \url{https://doi.org/10.1016/S0927-0507(05)80126-4}.

\bibitem[Miyazawa et~al.(2021)Miyazawa, Moura, Ota, and
  Wakabayashi]{phablo2021ejor}
F.~K. Miyazawa, P.~F. Moura, M.~J. Ota, and Y.~Wakabayashi.
\newblock Partitioning a graph into balanced connected classes: Formulations,
  separation and experiments.
\newblock \emph{European journal of operational research}, 293\penalty0
  (3):\penalty0 826--836, 2021.
\newblock URL \url{https://doi.org/10.1016/j.ejor.2020.12.059}.

\bibitem[Padberg and Rao(1982)]{PadbergRao1982}
M.~W. Padberg and M.~R. Rao.
\newblock Odd minimum cut-sets and b-matchings.
\newblock \emph{Mathematics of Operations Research}, 7\penalty0 (1):\penalty0
  67--80, 1982.
\newblock URL \url{https://doi.org/10.1287/moor.7.1.67}.

\bibitem[Rehfeldt et~al.(2022)Rehfeldt, Franz, and Koch]{rehfeldt2022optimal}
D.~Rehfeldt, H.~Franz, and T.~Koch.
\newblock Optimal connected subgraphs: Integer programming formulations and
  polyhedra.
\newblock \emph{Networks}, 80\penalty0 (3):\penalty0 314--332, 2022.
\newblock URL \url{https://doi.org/10.1002/net.22101}.

\bibitem[Samer(2023)]{samer2023}
P.~Samer.
\newblock On a class of strong valid inequalities for the connected matching
  polytope, 2023.
\newblock URL \url{https://doi.org/10.48550/arXiv.2309.14019}.
\newblock \textit{Submitted}.

\bibitem[Stockmeyer and Vazirani(1982)]{StockmeyerVazirani1982}
L.~J. Stockmeyer and V.~V. Vazirani.
\newblock {NP-completeness of some generalizations of the maximum matching
  problem}.
\newblock \emph{Information Processing Letters}, 15\penalty0 (1):\penalty0
  14--19, 1982.
\newblock URL \url{https://doi.org/10.1016/0020-0190(82)90077-1}.

\bibitem[Wang et~al.(2017)Wang, Buchanan, and Butenko]{wang2017imposing}
Y.~Wang, A.~Buchanan, and S.~Butenko.
\newblock On imposing connectivity constraints in integer programs.
\newblock \emph{Mathematical Programming}, 166:\penalty0 241--271, 2017.
\newblock URL \url{https://doi.org/10.1007/s10107-017-1117-8}.

\end{thebibliography}

\appendix
\clearpage


\section{Supplementary material}
\label{sec:appendix}

This supplement contains two long tables related to the experimental evaluation described in Section~\ref{sec:xp} in the paper.

Table~\ref{tab:instances} gives the cross-reference between numerical identifiers of each instance and their name in the original DIMACS implementation challenge benchmarks.

Table~\ref{tab:xp} contains the full numerical results of our implementation of both the compact extended formulation and the exponential one in the original space of variables.
On that table, LB corresponds to lower (primal) bounds, UB corresponds to upper (dual) bounds.
Column \textsf{\#Cuts} under the LP relaxation using $\mathcal{P}_{\text{full}}$ gives the number of inequalities added in our strengthened root node initialization (see Section~\ref{sec:bnc:general}).
In turn, the one for the ILP branch-and-cut scheme denotes the total number of user cuts and lazy constraints actually applied to improve the dual bound, retrieved from the solver final statistics.


\clearpage
\begin{landscape}
\begin{center}
\begin{ssmall}
\begin{longtable}{lllccclllcc}
\caption{Numerical IDs used for referencing DIMACs benchmark instances.}
\label{tab:instances}\\

\toprule
ID & Benchmark set & Instance name & $|V(G)|$ & $|E(G)|$          & \hspace{1cm} &     ID & Benchmark set & Instance name & $|V(G)|$ & $|E(G)|$ \\
\cmidrule{1-5}                                                                         \cmidrule{7-11}
\endfirsthead

\multicolumn{11}{c}
{{\tablename\ \thetable{} -- continued from previous page}} \\
\toprule
ID & Benchmark set & Instance name & $|V(G)|$ & $|E(G)|$          &&     ID & Benchmark set & Instance name & $|V(G)|$ & $|E(G)|$ \\
\cmidrule{1-5}                                                           \cmidrule{7-11}
\endhead

\multicolumn{11}{r}{{Continues on next page}} \\
\bottomrule
\endfoot

\bottomrule
\endlastfoot


1        &   MWCS-ACTMOD    &   drosophila001                    &     5226      &     93394                      &&             92       &   MWCS-JMPALMK   &   n-500-a-0.62-d-0.75-e-0.75              &     500       &     2597      \\        
2        &   MWCS-ACTMOD    &   drosophila005                    &     5226      &     93394                      &&             93       &   MWCS-JMPALMK   &   n-500-a-1-d-0.25-e-0.25                 &     500       &     6519      \\        
3        &   MWCS-ACTMOD    &   drosophila0075                   &     5226      &     93394                      &&             94       &   MWCS-JMPALMK   &   n-500-a-1-d-0.25-e-0.5                  &     500       &     6519      \\        
4        &   MWCS-ACTMOD    &   HCMV                             &     3863      &     29293                      &&             95       &   MWCS-JMPALMK   &   n-500-a-1-d-0.25-e-0.75                 &     500       &     6519      \\        
5        &   MWCS-ACTMOD    &   lymphoma                         &     2034      &     7756                       &&             96       &   MWCS-JMPALMK   &   n-500-a-1-d-0.5-e-0.25                  &     500       &     6519      \\        
6        &   MWCS-ACTMOD    &   metabol\_expr\_mice\_1           &     3523      &     4345                       &&             97       &   MWCS-JMPALMK   &   n-500-a-1-d-0.5-e-0.5                   &     500       &     6519      \\        
7        &   MWCS-ACTMOD    &   metabol\_expr\_mice\_2           &     3514      &     4332                       &&             98       &   MWCS-JMPALMK   &   n-500-a-1-d-0.5-e-0.75                  &     500       &     6519      \\        
8        &   MWCS-ACTMOD    &   metabol\_expr\_mice\_3           &     2853      &     3335                       &&             99       &   MWCS-JMPALMK   &   n-500-a-1-d-0.75-e-0.25                 &     500       &     6519      \\
\end{longtable}
\end{ssmall}
\end{center}
\end{landscape}

\clearpage
\begin{landscape}
\begin{center}
\begin{ssmall}
\begin{longtable}{lcccccccccccccccccccccc}
\caption{Complete numerical results of the extended formulation $(\ref{eq:compact:obj})$ using $\mathcal{P}_{\text{ext}}$ and the exponential one $(\ref{eq:full:obj})$ using $\mathcal{P}_{\text{full}}$, and their LP relaxations.}
\label{tab:xp}\\

\toprule
   &       &                   &&   \multicolumn{6}{c}{ LP relaxation }              &&     \multicolumn{12}{c}{ Integer linear programming } \\ 
                                    \cmidrule{5-10}                                         \cmidrule{12-23} 
\multicolumn{3}{c}{Instance}   &&   \multicolumn{2}{c}{ LP over $\mathcal{P}_{\text{ext}}$ } &&     \multicolumn{3}{c}{ LP over $\mathcal{P}_{\text{full}}$ }      &&    \multicolumn{5}{c}{ Black-box solver using $\mathcal{P}_{\text{ext}}$ } && \multicolumn{6}{c}{ Branch-and-cut scheme using $\mathcal{P}_{\text{full}}$ } \\
\cmidrule{1-3}                      \cmidrule{5-6} \cmidrule{8-10}                          \cmidrule{12-16} \cmidrule{18-23} 
ID & $|V|$ & $|E|$             &&   UB & Time (s) && UB & Time (s) & \#Cuts         &&     LB & UB & Gap (\%) & Time (s) & \#Nodes && LB & UB & Gap (\%) & Time (s) & \#Nodes & \#Cuts \\
\midrule 
\endfirsthead

\multicolumn{23}{c}
{{\tablename\ \thetable{} -- continued from previous page}} \\
\toprule
   &       &                   &&   \multicolumn{6}{c}{ LP relaxation }              &&     \multicolumn{12}{c}{ Integer linear programming } \\ 
                                    \cmidrule{5-10}                                         \cmidrule{12-23} 
\multicolumn{3}{c}{Instance}   &&   \multicolumn{2}{c}{ LP over $\mathcal{P}_{\text{ext}}$ } &&     \multicolumn{3}{c}{ LP over $\mathcal{P}_{\text{full}}$ }      &&    \multicolumn{5}{c}{ Black-box solver using $\mathcal{P}_{\text{ext}}$ } && \multicolumn{6}{c}{ Branch-and-cut scheme using $\mathcal{P}_{\text{full}}$ } \\
\cmidrule{1-3}                      \cmidrule{5-6} \cmidrule{8-10}                          \cmidrule{12-16} \cmidrule{18-23} 
ID & $|V|$ & $|E|$             &&   UB & Time (s) && UB & Time (s) & \#Cuts         &&     LB & UB & Gap (\%) & Time (s) & \#Nodes && LB & UB & Gap (\%) & Time (s) & \#Nodes & \#Cuts \\
\midrule 
\endhead

\multicolumn{23}{r}{{Continues on next page}} \\
\bottomrule
\endfoot

\bottomrule
\endlastfoot


1        &     5226      &     93394     &&                 51.13     &     34.06        &&       28.49     &     300.06    &    729                    &&                     18.75     &     41.87     &     123.33    &     3600.27   &     1                 &&             20.32     &     20.32     &     0.00      &     1583.24   &     1952      &         330               \\     
2        &     5226      &     93394     &&                 197.54    &     180.67       &&       165.13    &     300.46    &    489                    &&                     155.48    &     179.37    &     15.37     &     3600.23   &     1                 &&             155.48    &     155.48    &     0.00      &     2015.71   &     1780      &         257               \\     
3        &     5226      &     93394     &&                 278.08    &     109.50       &&       245.28    &     300.77    &    382                    &&                     233.19    &     267.85    &     14.86     &     3600.23   &     1                 &&              0.00     &     236.24    &      --       &     3300.09   &     1437      &         277               \\     
4        &     3863      &     29293     &&                 8.42      &     2.97         &&       4.61      &     12.46     &    150                    &&                     4.36      &     4.36      &     0.00      &     82.46     &     1812              &&             4.36      &     4.36      &     0.00      &     0.51      &     1         &         0                 \\     
5        &     2034      &     7756      &&                 74.62     &     0.85         &&       60.36     &     7.08      &    238                    &&                     60.36     &     60.36     &     0.00      &     49.43     &     1                 &&             60.36     &     60.36     &     0.00      &     0.19      &     1         &         1                 \\
\end{longtable}
\end{ssmall}
\end{center}
\end{landscape}


\end{document}